\documentclass[intlimits,twoside,a4paper]{article}
\usepackage{amsmath,amssymb}
\usepackage{graphicx}
\usepackage{wrapfig}
\usepackage[T2A]{fontenc}
\usepackage[cp1251]{inputenc}

\usepackage[eqsecnum]{cmpj2}
\issue{2013}{16}{4}{43601}
\doinumber{10.5488/CMP.16.43601}

\title[Selective adsorption in slit-systems]%
{Selective adsorption of ions in charged slit-systems%
\thanks{We dedicate this paper to Myroslav Holovko in honour of his valuable contribution to the field of statistical mechanics of fluids.}
}
\author[M.~Valisk\'o, D.~Henderson, D.~Boda]{M.~Valisk\'o\refaddr{label1}\thanks{E-mail: valisko@almos.vein.hu}\,, D.~Henderson\refaddr{label2}, D.~Boda\refaddr{label1,label2}
}

\authorcopyright{M.~Valisk\'o, D.~Henderson, D.~Boda, 2013}
\addresses{
 \addr{label1} Department of Physical Chemistry, University of Pannonia, P. O. Box 158, Veszpr\'em, Hungary
 \addr{label2} Department of Chemistry and Biochemistry, Brigham Young University, Provo, UT, USA
}

\date{Received July 27, 2013, in final form August 6, 2013}

\begin{document}

\maketitle

\begin{abstract}
We study the selective adsorption of various cations into a layered slit system using grand canonical Monte Carlo simulations.
The slit system is formed by a series of negatively charged membranes.
The electrolyte contains two kinds of cations with different sizes and valences modelled by charged hard spheres immersed in a continuum dielectric solvent.
We present the results for various cases depending on the combinations of the properties of the competing cations.
We concentrate on the case when the divalent cations are larger than the monovalent cations.
In this case, size and charge have counterbalancing effects, which results in interesting selectivity phenomena.

\pacs 61.20.Qg, 68.03.-g, 81.05.Rm, 61.20.Ja, 07.05.Tp
\end{abstract}

\section{Introduction}
\label{sec:intro}

This work was motivated by two previous papers \cite{valisko-jpcc-111-15575-2007,kovacs-cmp-15-23803-2012}, and, also, it is a direct continuation of those works.
Ion selectivity was a central theme of many of our works for ion channels \cite{boda-jpcb-104-8903-2000,boda-jpcb-105-11574-2001,boda-mp-100-2361-2002,boda-pccp-4-5154-2002,boda-ms-30-89-2004,boda-jcp-125-034901-2006,boda-prl-98-168102-2007,boda-bj-93-1960-2007,gillespie-bj-95-2658-2008,boda-jgp-133-497-2009,rutkai-jpcl-1-2179-2010,malasics-bba-1798-2013-2010,boda-jcp-134-055102-2011,csanyi-bba-1818-592-2012,boda-jml-inpress-2013} which, in turn, motivated our study for the selective adsorption of various ions at highly charged interfaces \cite{valisko-jpcc-111-15575-2007}.
In that work, we simulated --- using Monte Carlo (MC) --- the adsorption of cations at a charged electrode, where the size and charge of cations were varied.
We considered the competition of small monovalent (Sm C$^{+}$), small divalent (Sm C$^{2+}$), large monovalent (Lg C$^{+}$), and large divalent (Lg C$^{2+}$) ions.
We were interested in the question which species is adsorbed in larger quantity in the double layer (DL) in competition with other species for various electrode charges and mole fractions.

This is the question that we pose in this work too, except that the ions now are adsorbed in slits instead of near an isolated wall.
As a matter of fact, double layers are commonly simulated between two confining walls, where either of them can be charged and serve as an electrode.
If the distance of the walls is large enough, the DLs formed at the walls are independent of each other and a charge neutral bulk region is formed in the middle of the simulation cell.
The reference point for the electrical potential then can be set in this bulk region.

If the walls, however, are close to each other, the DLs overlap, and the bulk region disappears.
In this case, we talk about a slit.
Slits are generally simulated in the grand canonical (GC) ensemble, where the electrolyte in the slit is in equilibrium with a virtual bulk phase represented by its temperature and the chemical potentials of the ionic species.
The bulk system is a virtual bath where the ions go and where they come from during the insertion/deletion steps of the Grand Canonical Monte Carlo (GCMC) simulations.
In this case, however, the ground of the electrical potential cannot be set in this bulk phase, because it is not present in the simulation cell and Poisson's equation cannot be integrated over it.
The ground of the potential, therefore, is ill defined.
Many studies for electrolytes confined in a slit considered only a single slit in equilibrium with a bulk in the GC ensemble \cite{lozadacassou_jcp_1990,pizio04,patryk04,pizio05c,yu06,borowko05,pizio05a,buyukdagli_jsm_2011,ibarraarmenta_pccp_2011,martinm_jpcb_113_2009,martinmolina_sm_2010,wang_jcp_2011,pizio_jcp_2012,pizio_jcp_2013,sanchezarellano_jml_2013}.
The behavior of fluids near surfaces attracted the attention of many workers \cite{holovko_cmp_2007,patsahan_cmp_2009,dicaprio_jpcb_2009}.
Electrolytes are especially interesting due to the forming DLs.
If the wall is charged, well-defined contact value theorems describe the quantitative behavior of ions at the electrode \cite{henderson-jcp-75-2025-1981,holovko_jcp_2005,bhuiyan_jeac_2007,holovko_jcp_2007,holovko_jcp_2008}.

In our previous work \cite{kovacs-cmp-15-23803-2012}, therefore, we simulated a slit system, where we allowed the existence of several slits and, more importantly, the existence of two baths on the two sides of the slit system.
The slit system was composed of membranes of a fixed width carrying surface charges.
The slits between the membranes then adsorbed cations in order to balance the membrane charges.
Our main interest in that study \cite{kovacs-cmp-15-23803-2012} was the electrical conditions present inside and outside the slit system, namely, the profiles for the charge density, the electric field, and the  mean electrical potential.
The electrical potential difference between the inner slits and the bath had two main components.
First, DLs were formed outside the slit system.
The potential drop associated with these DLs formed the main component of the total drop.
Additionally, there was a potential drop across the outer slits.
We analyzed the effect of the number of slits, the width of the slits, membrane charge, concentration, and the presence of divalent ions.
The electrolyte, however, was a pure electrolyte meaning that only one kind of cation and one kind of anion were present.

In this paper, we unite the scopes of the two previous works \cite{valisko-jpcc-111-15575-2007,kovacs-cmp-15-23803-2012} and study the competitive adsorption of two different cations in slit systems.
Slit systems are first-order models for layered silicate minerals \cite{deville-langmuir-1998,jonsson-2004,jonsson-2005,pegado_jpcm_2008}, porous electrodes \cite{kiyohara07,kiyohara_jcp_2010,kiyohara_jcp_2011,KiyoharaJPCC,kiyohara_smc_2011}, and lyotropic lamellar liquid crystals \cite{jonsson_jpc_1980,ekwall}.
The structure, swelling, and adsorption properties of such materials (e.g., kaolinite, montmorillonite) are subject of extensive experimental and simulation studies \cite{shroll_jcp_1999,chavez-paez,rutkai_cpl_2008,rutkai_jcis_2009,mako_jcis_2010}.
Collections of slits can also be considered as simple models of  porous media.
The adsorption of various fluids in porous matrices was extensively studied by theoretical and simulation methods \cite{vlachy_jacs_1989,vlachy_ajc_1990,vlachy_jeac_1990,jamnik_jacs_1993,jamnik_jacs_1995,trokhymchuk_jpc_1996,trokhymchuk_jcp_1997,hribar_jcp_1997,hribar_jcp_1998,hribar_jpcb_1999,hribar_jpcb_2000,hribar_jpcb_2001,vlachy_jpcb_2004,luksic_cmp_2012}.
% In this work, we focus on the electrical properties of charged slit systems and show concentration, charge, electric field, and potential profiles for different geometrical parameters (width of the slit, width of the membrane), membrane charge, and electrolytes (concentration, ionic charge).

\section{Model and method}
\label{sec:model}

The electrolyte is modelled by the Primitive Model (PM).
In this model, the solvent is represented by its dielectric response characterized by the dielectric constant $\epsilon$, while the ions are represented by charged hard spheres interacting through the screened Coulomb + hard sphere pair potential:
\begin{equation}
u(r_{ij}) =
\left\lbrace
\begin{array}{ll}
\infty & \quad \mathrm{for} \quad r_{ij}<R_{i}+R_{j}\,, \\
 \dfrac{1}{4\pi\epsilon_{0}\epsilon} \dfrac{q_{i}q_{j}}{r_{ij}} & \quad \mathrm{for} \quad r_{ij} \geqslant  R_{i}+R_{j}\,,\\
\end{array}
\right.
\label{eq:uij}
\end{equation}
where $r_{ij}$ is the distance of two ions, $q_i$ is the ionic charge ($ q_{i}=z_{i}e$, $z_{i}$ being the valence and $e$ the elementary charge), $R_i$ is the ionic radius, and $\epsilon_0$ is the vacuum's permittivity.
% In the RPM, $R_{i}$ is the same for every ionic species ($R=1.5$~{\AA} in this work).
We use $R_{\mathrm{A}}=1$~{\AA} for the radius of the anions, while we use $R_{\mathrm{Sm\, C}}=1$~{\AA} and  $R_{\mathrm{Lg\, C}}=2.1$~{\AA} for the radii of the small and large cations, respectively.
The ionic charges are point charges in the centers of hard spheres.

The $\alpha$th membrane is confined by two hard walls at $x_{\alpha}^{\mathrm{L}}$ and $x_{\alpha}^{\mathrm{R}}$, where each hard wall can carry a $\sigma_{\mathrm{M}} $ surface charge.
The interaction potential between such a charged hard wall and an ion is
\begin{equation}
v_{i}(|x|)=
\left\lbrace
\begin{array}{ll}
\infty & \quad \mathrm{for} \quad |x|<R_{i}\,, \\
 -\dfrac{z_{i}e\sigma_{\mathrm{M}}|x|}{2\epsilon_{0}\epsilon} & \quad \mathrm{for} \quad |x| \geqslant  R_{i}\,,
\end{array}
\right.
\label{eq:vi}
\end{equation}
where $|x|$ is the distance of the ion from the surface.

There are $N_{\mathrm{M}}$ membranes of width $L_{\mathrm{M}}$ (the distance of the two walls forming the membrane: $L_{\mathrm{M}}=x_{\alpha}^{\mathrm{R}}-x_{\alpha}^{\mathrm{L}}$ for every $\alpha$) in the slit system.
The value $L_{\mathrm{M}}=10$~{\AA} was used throughout this work.
In this work, both membrane walls carry $\sigma_{\mathrm{M}}$  surface charge.
The distance of two membranes, namely, the width of the slit is $L_{\mathrm{S}}$.
This distance is kept fixed during the simulation, namely, the slit system is rigid.
Ions are not allowed to enter the membranes, so the membrane surfaces behave as hard walls.
Obviously, the number of slits is $N_{\mathrm{S}}=N_{\mathrm{M}}-1$.
There are two bulk phases of widths $L_{\mathrm{B}}$ on the two sides of the slit system.
The simulation cell is closed by two hard walls on the left hand side of the left hand bulk region and on the right hand side of the right hand  bulk region.
In this study, these walls are uncharged.

GCMC simulations have been performed for the system described above.
Periodic boundary conditions have been applied in the $y-z$ dimensions.
The effect of the ions outside the central simulation cell was taken into account by the charged sheet method proposed by Torrie and Valleau \cite{torrie-jcp-1980} and developed further by Boda et al. \cite{boda-jcp-109-7362-1998}.

In GCMC simulations of the DL, in addition to the usual particle displacement steps, we insert and delete neutral clusters of ions, e. g., $\nu_{+}$ cations and $\nu_{-}$ anions ($\nu_{+}$ and $\nu_{-}$ being the stoichiometric coefficients).
This way, we assure that the simulation cell is charge neutral in every instant of the simulation.
The acceptance probability of these steps is found in our previous paper \cite{kovacs-cmp-15-23803-2012}.
The chemical potentials needed for the GCMC simulations' input were determined by the Adaptive-GCMC method of Malasics et al. \cite{malasics-jcp-128-124102-2008,malasics-jcp-132-244103-2010}.

The dimensions of the simulation cell in the $y-z$ dimensions is in the range of 120--150~{\AA}.
System-size checks indicated little sensitivity of the concentration profiles on the $y-z$ dimensions of the cell.
In a typical simulation, the sample contained several hundreds of millions ($10^{8}$) configurations.

The main output quantities of the simulations are the density profiles of various ionic species, $\rho_{i}(x)$, from which the ionic charge profile is obtained as follows:
\begin{equation}
 q(x) = \sum_{i} z_{i}e \rho_{i}(x).
\label{eq:qdef}
\end{equation}
Poisson's equation
\begin{equation}
\dfrac{\rd^{2}\Phi (x)}{\rd x^{2}}= - \dfrac{1}{\epsilon_{0}\epsilon} q_{\mathrm{tot}}(x)
\label{poisson}
\end{equation}
is solved for the mean electrical potential $\Phi(x)$ and the electric field $E(x)=-\rd\Phi(x)/\rd x$ using Neumann boundary conditions.
The total charge density
\begin{equation}
 q_{\mathrm{tot}}(x) = q(x) + \sum_{\mathrm{M}} \sigma_{i} \delta (x-x_i) ,
\end{equation}
that contains all charges in the system [including the $\sigma_{\mathrm{M}}$ membrane charges in addition to the ionic charge density $q(x)$] is integrated once to get $E(x)$ and once more to get $\Phi(x)$ with the boundary condition $E(-\infty)=0$.
The system is always charge neutral due to the design of the GCMC insertion/deletion steps.
The other condition $E(\infty)=0$, therefore, is automatically satisfied.
The zero level of the potential can be chosen freely.
The equations are found in \cite{kovacs-cmp-15-23803-2012}.

\section{Results and discussion}
\label{sec:res}

In this work, we measure distances in~{\AA}, so particle densities are measured in~{\AA}$^{-3}$.
In the figures, however, we plot concentration profiles that are related to the density profiles through $c_{i}(x)= 1660.58 \cdot  \rho_{i}(x)$.
The charge profile is also computed in terms of concentrations and is normalized by the elementary charge: $q(x)/e=\sum_{i}z_{i}c_{i}(x)$ (the unit is M, which is mol/dm$^{3}$).
The potential is plotted in a dimensionless form as $e\Phi(x)/k_{\mathrm{B}}T$.
The electric field is the derivative of this dimensionless potential, so its unit is~{\AA}$^{-1}$.
We will denote it by $E^{*}(x)$.

The temperature is $T=298.15$ K.
We show the results for $N_{\mathrm{S}}=3$ slits; other numbers gave similar results.
We have several variables that can be changed: valences, radii, and concentrations of cations, width of slits ($L_{\mathrm{S}}$), and membrane charge ($\sigma_{\mathrm{M}}$).
We used slit widths $L_{\mathrm{S}}=5$, 10, and 20~{\AA}, but most results will be shown for 5~{\AA}.
We changed the membrane charge in the interval $-0.3\leqq \sigma_{\mathrm{M}}\leqq 0$~C\,m$^{-2}$.
We change ion concentrations in a way that the ionic strength
\begin{equation}
I=\dfrac{1}{2}\sum_{i=1}^{3}z_{i}^{2}c_{i}
\end{equation}
is constant in a series of calculations.
The composition of the mixture can be expressed in terms of either mole fraction
\begin{equation}
x_{i}^{\mathrm{c}}=\dfrac{[\mathrm{C}_{i}]}{[\mathrm{C}_{1}]+[\mathrm{C}_{2}]}
\end{equation}
or in terms of a fraction of ionic strength
\begin{equation}
x_{i}^{\mathrm{I}}=\dfrac{I_{i}}{I}\,,
\end{equation}
where $I_{i}=\frac{1}{2}(z_{i}^{2}+z_{i})[\mathrm{C}^{z_{i}}]$ and $I=I_{1}+I_{2}$, where it was assumed that the anion is monovalent.
If species $i$ is monovalent, $I_{i}=[\mathrm{C}^{z_{i}}]$.
If species $i$ is divalent, $I_{i}=3[\mathrm{C}^{z_{i}}]$.
If the two cations have the same valence, the two fractions are the same.
If one of the species is monovalent and the other is divalent, they can be transformed to each other through
\begin{equation}
x_{i}^{\mathrm{c}}=\dfrac{3x_{i}^{\mathrm{I}}}{2x_{i}^{\mathrm{I}}+1}\,.
\end{equation}
We will plot our results as functions of $x_{i}^{\mathrm{I}}$, but we must be aware of the difference from the classical mole fraction.
\begin{figure}[ht]
\includegraphics[width=0.48\textwidth]{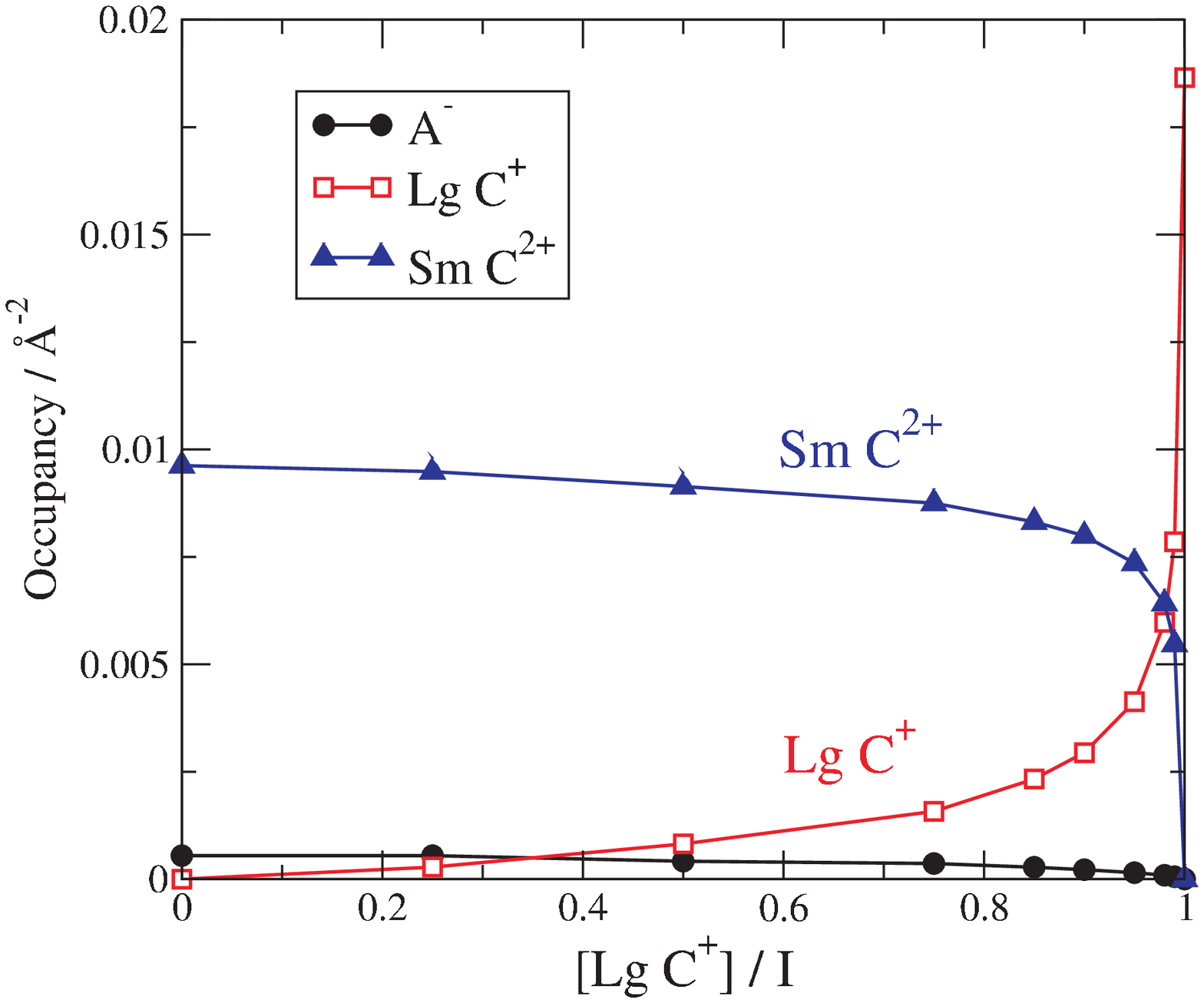}%
\hfill%
\includegraphics[width=0.48\textwidth]{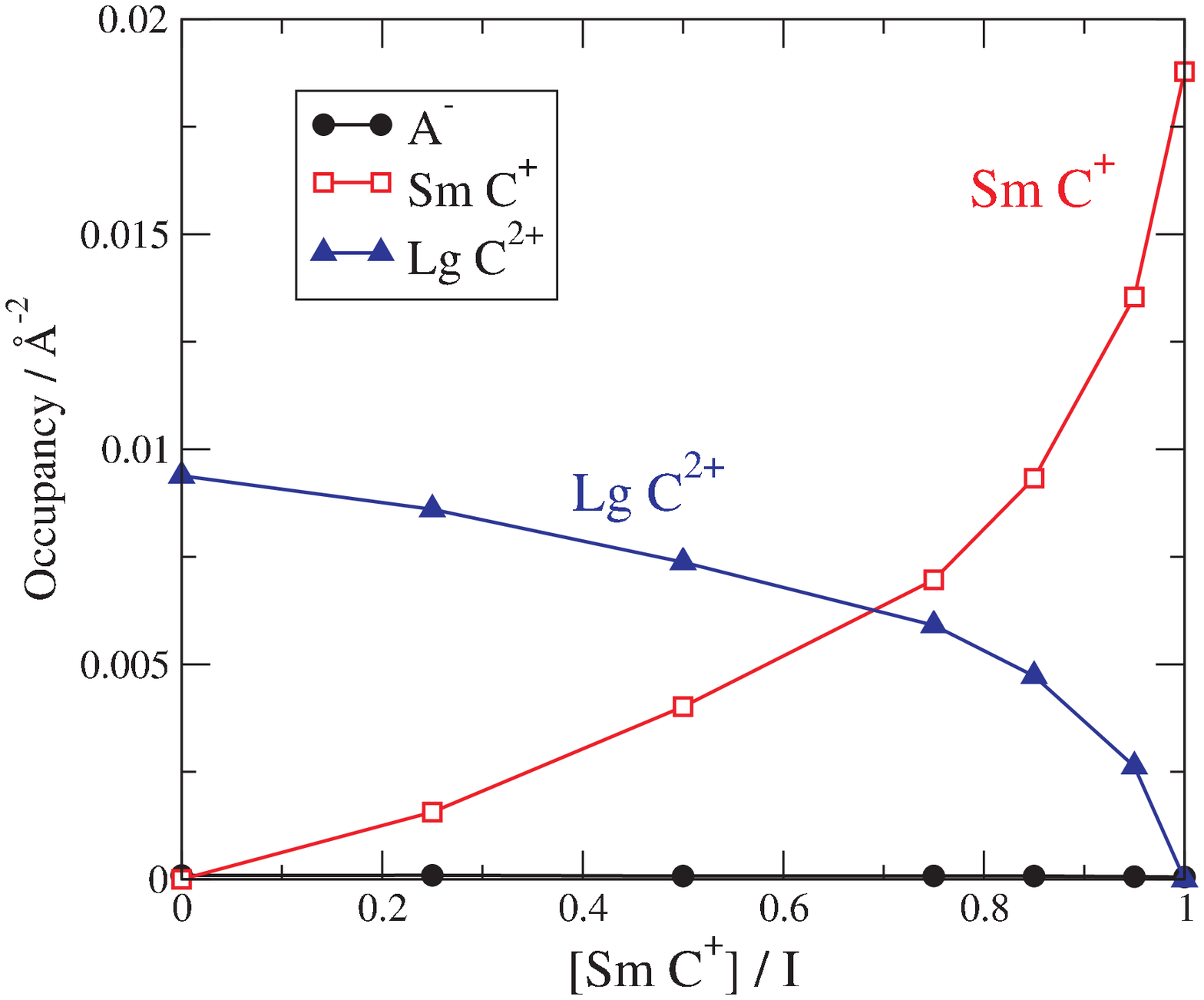}%
\\%
\parbox[t]{0.48\textwidth}{%
\centerline{(a)}%
}%
\hfill%
\parbox[t]{0.48\textwidth}{%
\centerline{(b)}%
}%
\\
\includegraphics[width=0.48\textwidth]{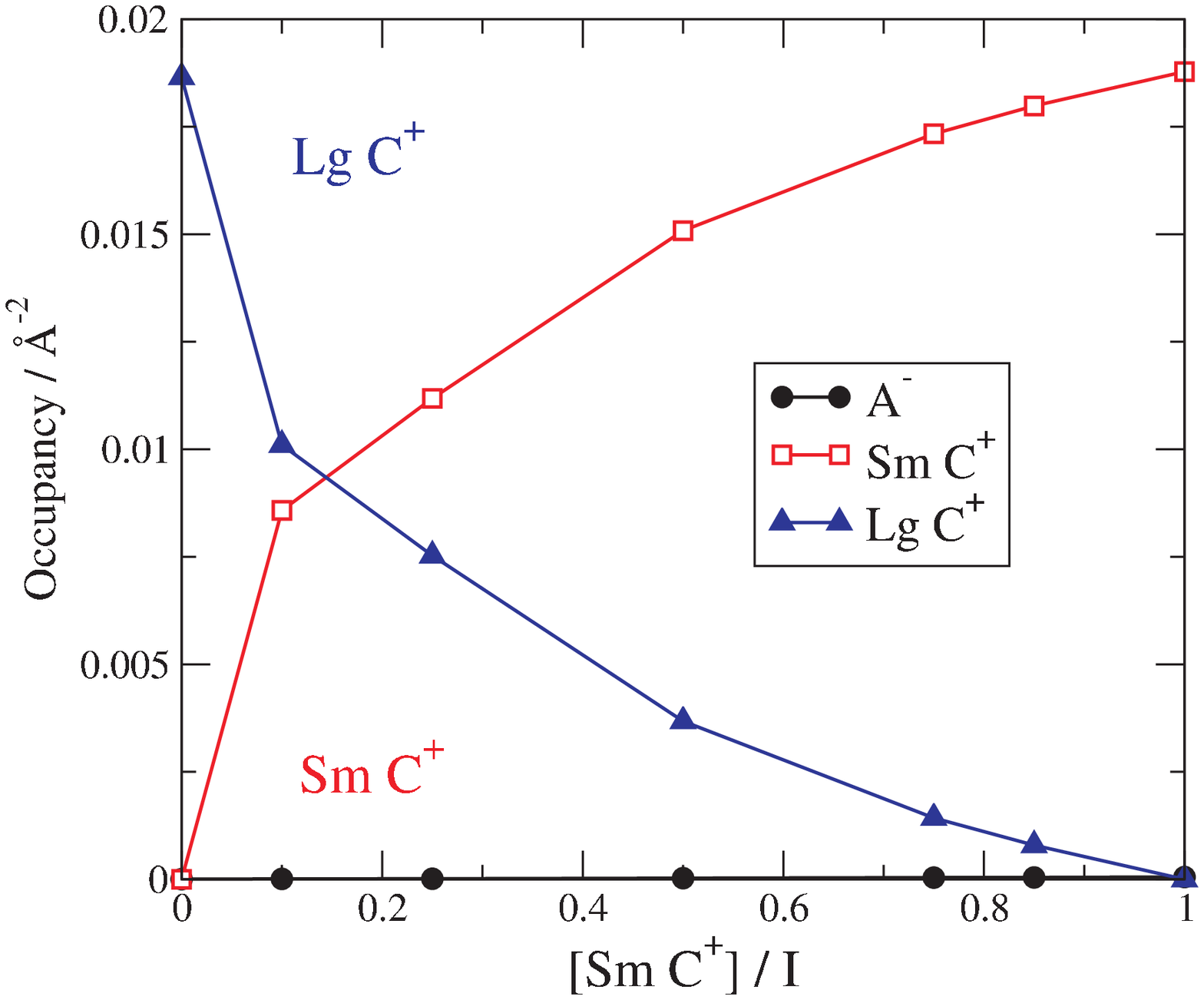}%
\hfill%
\includegraphics[width=0.48\textwidth]{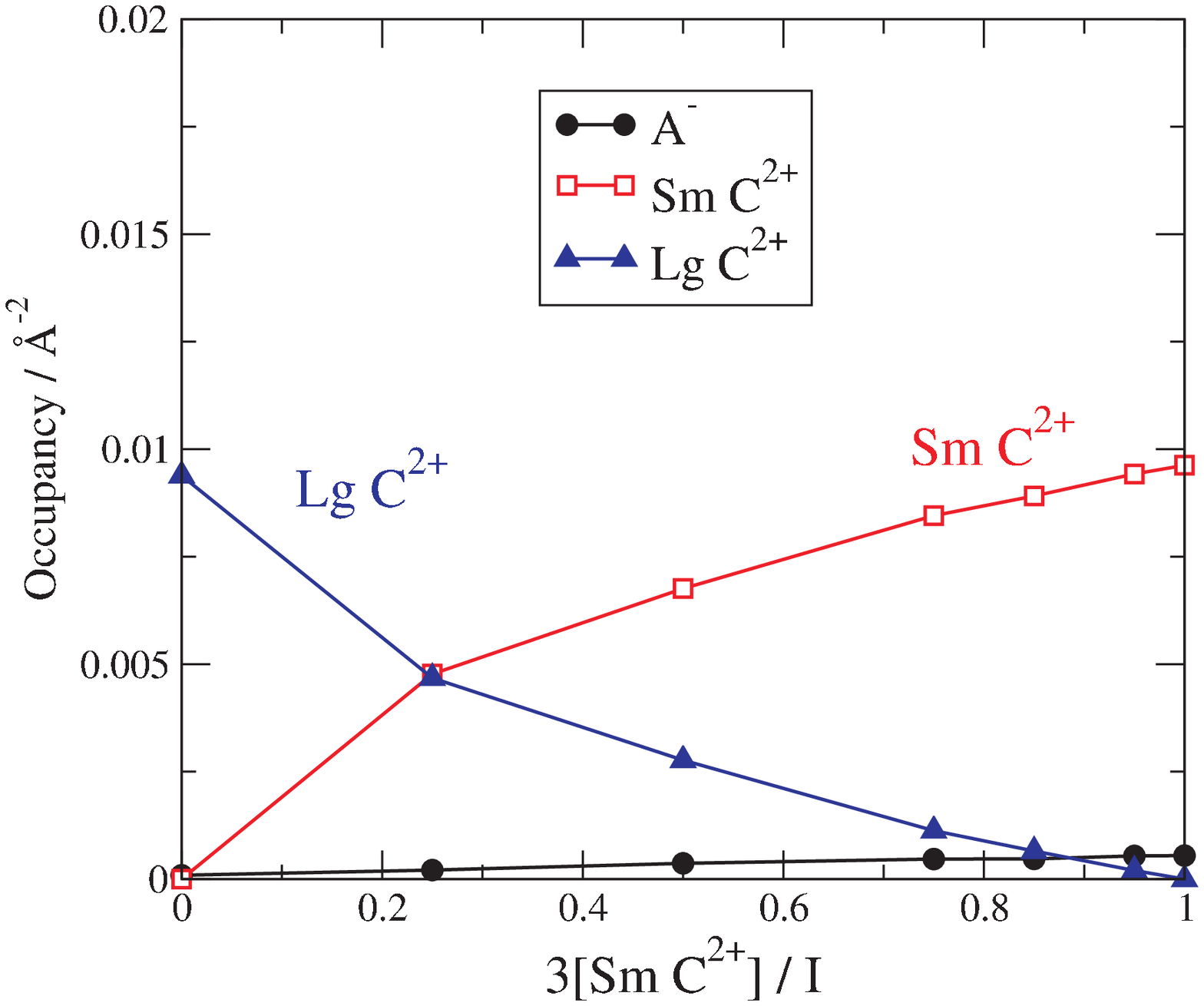}%
\\%
\parbox[t]{0.48\textwidth}{%
\centerline{(c)}%
}%
\hfill%
\parbox[t]{0.48\textwidth}{%
\centerline{(d)}%
}%
\caption{(Color online) Competition between various cations in the central slit ($N_{\mathrm{S}}=3$) for membrane charge $\sigma_{\mathrm{M}}=-0.15$~C\,m$^{-2}$.
Occupancies are shown as functions of the ionic-strength mole fraction of one of the competing species.
Mole fraction expressed in terms of ionic strength is $[\mathrm{Sm\, C}^{+}]/I$.
While varying mole fraction, the ionic strength of the system was kept constant at $I=0.4$ M.}
\label{fig1}
\end{figure}

To characterize the selectivity behavior of the slits, we show selectivity (or binding) curves, in which we plot occupancies against a quantity characterizing composition, $x_{1}^{\mathrm{I}}$, in this work.
We define occupancy as the average number of an ionic species in the central slit divided by the cross section of the simulation cell.
Occupancy, therefore, is a surface density; its unit is $\mbox{\AA}^{-2}$.
There are other ways to characterize selectivity, however.
We can plot ion exchange isotherms, where the mole fraction of one of the competing species inside the slit is plotted as a function of the mole fraction of the same species in the bulk.
Jamnik and Vlachy \cite{jamnik_jacs_1995} showed the results for the separation factor
\begin{equation}
K=\dfrac{c_{2}^{\mathrm{pore}}/c_{1}^{\mathrm{pore}}}{c_{2}^{\mathrm{bulk}}/c_{1}^{\mathrm{bulk}}}\,,
\end{equation}
where $c_{i}^{\mathrm{pore}}$ and $c_{i}^{\mathrm{bulk}}$ are ion concentrations in the pore and the bulk.
These parameters are used in ion exchange resins and other applications, where separation of components is the main goal.
Selectivity is the basis of these processes.
While the ion exchange isotherm and the separation factor have the advantage that they characterize selectivity with only one curve, showing the occupancies of every ionic species gives more information on the behavior of the system.

Figure~\ref{fig1} shows selectivity curves for four different cases.
The membrane charge is $\sigma_{\mathrm{M}}=-0.15$~C\,m$^{-2}$ and the slit width is $L_{\mathrm{S}}=5$~{\AA} in every case.
As in our previous study \cite{valisko-jpcc-111-15575-2007}, we consider the following competitions: (a) Lg C$^{+}$ vs. Sm C$^{2+}$, (b)  Sm C$^{+}$ vs. Lg C$^{2+}$, (c)
Sm C$^{+}$ vs. Lg C$^{+}$, and (d)  Sm C$^{2+}$ vs. Lg C$^{2+}$.

In agreement with our ion channel studies, the slit is selective for the smaller ion as figures~\ref{fig1}~(c)--(d) show.
This is hardly a surprise.
In the other two cases, monovalent ions compete with divalent ions.
The case shown in figure~\ref{fig1}~(a) was called the ``selective'' case, where the small divalent has the advantages of both the larger charge and the smaller size.
The other case of figure~\ref{fig1}~(b) was called the ``competitive'' case, where large divalent ions compete with small monovalent ions.
In this case, the smaller size is an advantage for the monovalents, while the larger charge is an advantage for the divalents.
We will focus on this case hereinafter.
\begin{figure}[!h]
\includegraphics[width=0.48\textwidth]{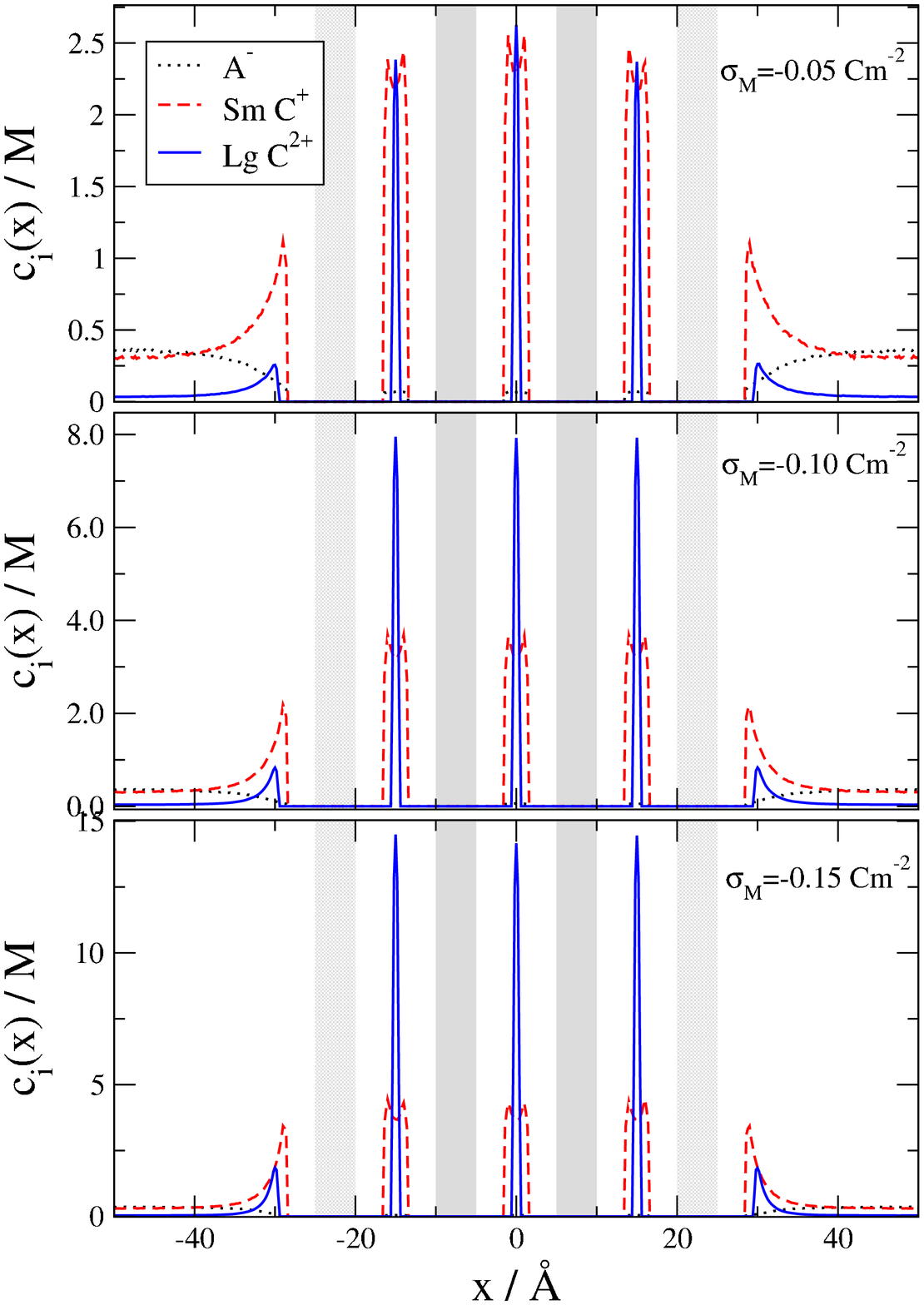}%
\hfill%
\includegraphics[width=0.48\textwidth]{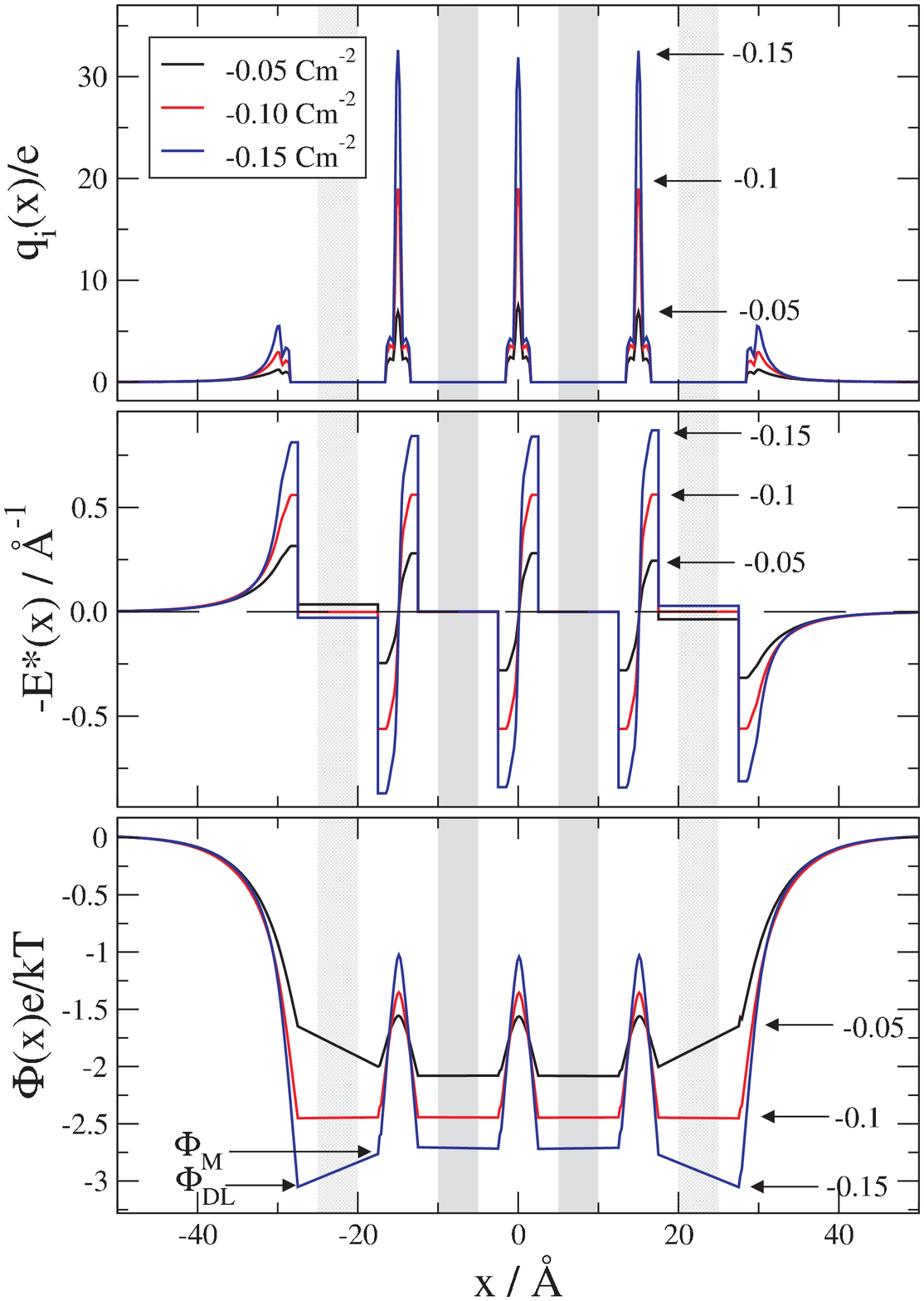}%
\\%
\parbox[t]{0.48\textwidth}{%
\centerline{(a)}%
}%
\hfill%
\parbox[t]{0.48\textwidth}{%
\centerline{(b)}%
}%
\caption{(Color online) (a) Concentration profiles of various ions for the ``competitive'' case (Sm C$^{+}$ vs. Lg C$^{2+}$).
The various panels refer to various membrane charges.
Bulk concentrations are $[\mathrm{Sm\, C}^{+}]=0.3$ M and $[\mathrm{Lg\, C}^{2+}]=0.0333$ M with $I=0.4$ total ionic strength ($[\mathrm{Sm\, C}^{+}]/I=0.75$).
Slit width is $L_{\mathrm{S}}=5$~{\AA}.
(b) The charge, electric field, and potential profiles for these cases. The unit of the reduced electric field is~{\AA}$^{-1}$.
The arrows and numbers near curves indicate the value of the membrane charge.
}
\label{fig2}
\end{figure}

Figure~\ref{fig2} shows various profiles for the ``competitive'' case for different membrane charges for a fixed composition ($x_{\mathrm{Sm\, C}^{+}}^{\mathrm{I}}=0.75$).
Panel (a) shows that the concentrations of both ions increase in the slits as $|\sigma_{\mathrm{M}}|$ is increased, but the divalent increases more.
Panel (b) shows the charge, electric field, and potential profiles.
The electric field profiles show that the two outer slits are charged.
The sign of the charge, however, depends on the membrane charge.

For $\sigma_{\mathrm{M}}=-0.1$~C\,m$^{-2}$, $-E^{*}(x)$ is zero in the left hand side membrane.
This is the case when the slits are charge neutral (together with the membrane charges on the confining walls) and the charge of the outer DL exactly balances the surface charges on the outer walls of the slit system.
The potentials are flat in the membranes in these cases, because the membrane charges are always balanced by the surrounding cations.

For $\sigma_{\mathrm{M}}=-0.05$~C\;m$^{-2}$, $-E^{*}(x)$ is positive in the left hand side membrane.
Since $-E^{*}(x)$ is the integral of the charge profile, it means that there is more cationic charge in the outer DLs and less in the outer slits compared to the $\sigma_{\mathrm{M}}=-0.1$~C\,m$^{-2}$ case, namely, we ``overcharge'' the DLs.
This case results in a potential profile also observed in our previous work \cite{kovacs-cmp-15-23803-2012}; the potential drop is smaller (in absolute value) in the outer DLs than in the central membranes.
If we denote the potential drop in the DL by $\Phi_{\mathrm{DL}}$ and the potential drop in the central membranes by $\Phi_{\mathrm{M}}$ [called membrane potential, see figure~\ref{fig2}~(b)], then $|\Phi_{\mathrm{DL}}|<|\Phi_{\mathrm{M}}|$.
% The membrane potential (multiplied by the charge) can be viewed as the mean-field component of the excess chemical potential of an ion.

For  $\sigma_{\mathrm{M}}=-0.15$~C\,m$^{-2}$, the situation is the opposite, we ``overcharge'' the outer slits.
In this case,  $|\Phi_{\mathrm{DL}}|>|\Phi_{\mathrm{M}}|$, which results in a peculiar behavior of the potential; the potential is deeper in the outer DL than in the inner membranes.

An interpretation of the potential profile can be given as follows.
Gillespie suggested breaking the excess chemical potential into various components (hard sphere, electrical, mean field, screening) in his Density Functional theoretical study of the Ryanodine Receptor calcium channel \cite{Dirkbj,gillespie_bj_amfe_2009}.
This made an energetic analyzis of ion selectivity in the selectivity filter (acting as a binding site) of calcium channels possible.
In a GCMC study \cite{boda-jcp-134-055102-2011}, we extended this approach to the presence of inhomogeneous dielectric.
It was also used to analyze selective adsorption at highly charged interfaces \cite{valisko-jpcc-111-15575-2007}.
The mean-field term appearing in this analysis is the interaction of the ion with the mean potential, $z_{i}e\Phi(x)$.
This term provides information on how the ion interacts with the average electrical potential.
The negative well of the mean electrical potential means that the divalent cation has an electrical advantage by interacting with the negative mean potential.
There are, however, other terms that describe hard sphere exclusion or ionic correlations.
These terms counterbalance the mean field term and make the chemical potential constant.
All these terms together make it possible for the ions to be in equilibrium with the bath.
The mean electrical potential is only one term of the many.

\begin{figure}[b]
\includegraphics[width=0.32\textwidth]{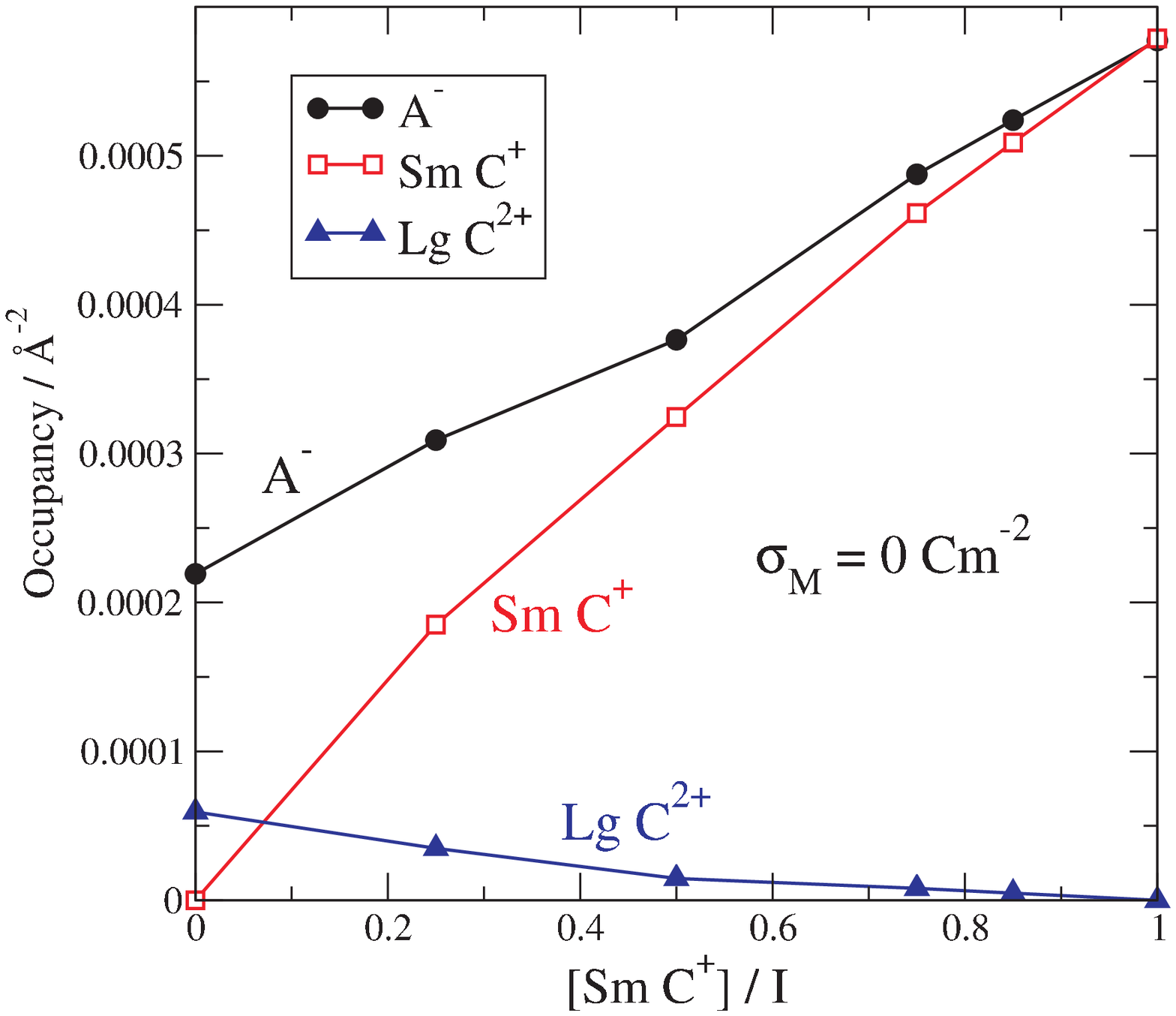}%
\hfill%
\includegraphics[width=0.32\textwidth]{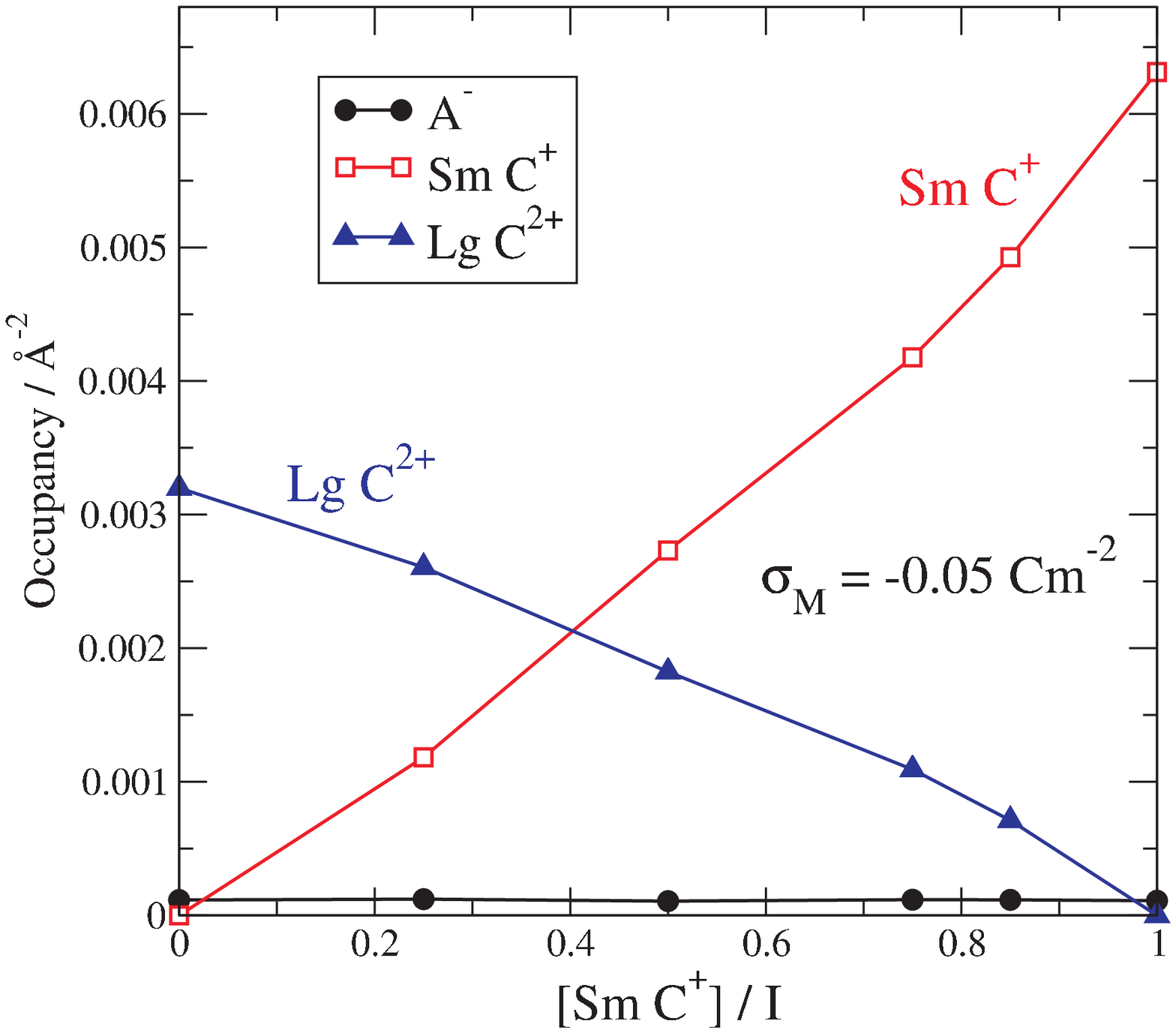}%
\hfill%
\includegraphics[width=0.32\textwidth]{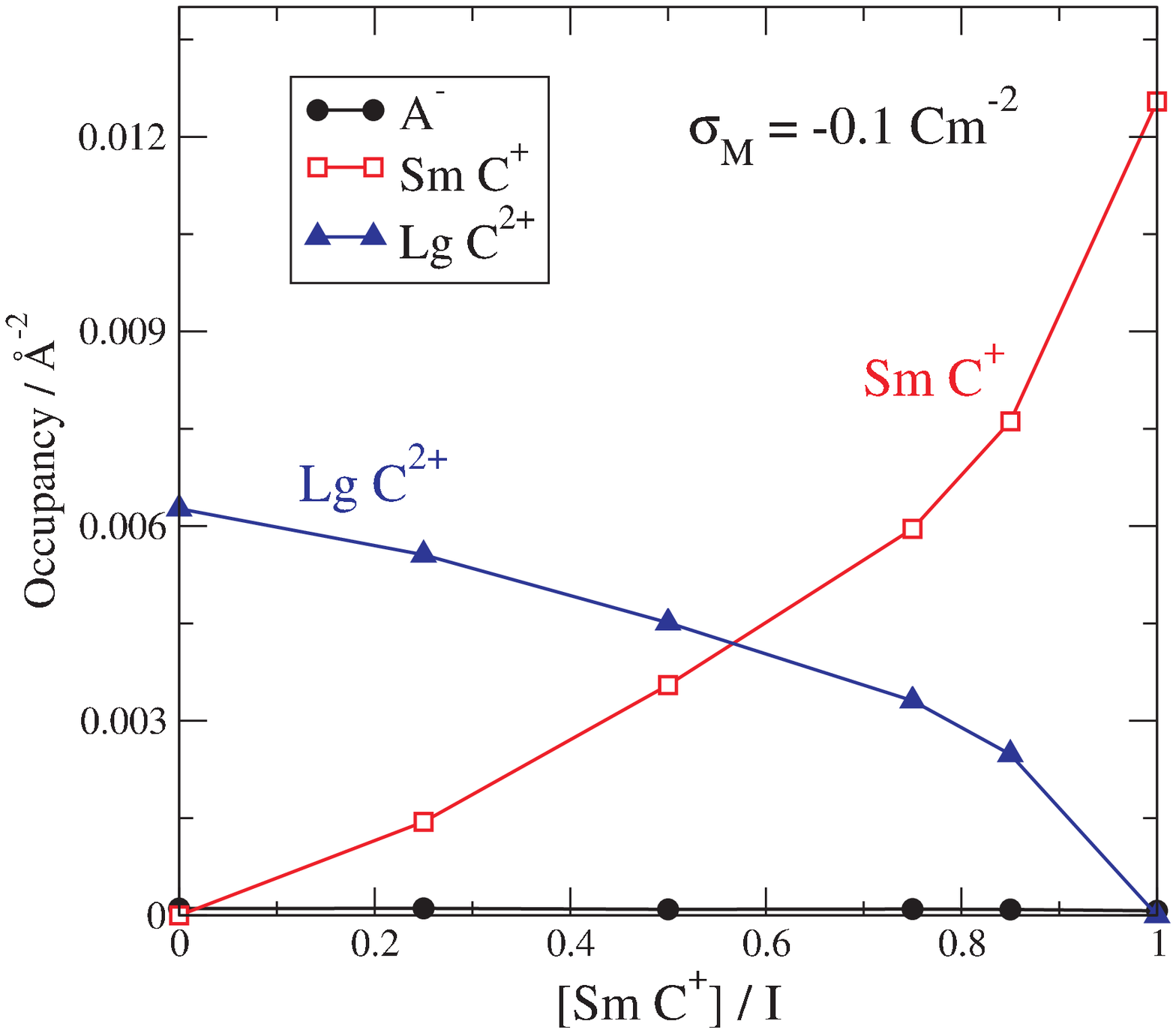}%
\\%
\parbox[t]{0.32\textwidth}{%
\centerline{(a)}%
}%
\hfill%
\parbox[t]{0.32\textwidth}{%
\centerline{(b)}%
}%
\hfill%
\parbox[t]{0.32\textwidth}{%
\centerline{(c)}%
}%
\caption{(Color online) Competition between small monovalent and large divalent cations (``competitive'' case) in the central slit ($N_{\mathrm{S}}=3$) for various membrane charges $\sigma_{\mathrm{M}}=0$, $-0.05$, and $-0.1$~C\,m$^{-2}$.
Occupancies are shown as functions of $[\mathrm{Sm\, C}^{+}]/I$ with $I=0.4$~M.
}
\label{fig3}
\end{figure}

Figure~\ref{fig1} shows the selectivity curves for a specific membrane charge, $\sigma_{\mathrm{M}}=-0.15$~C\,m$^{-2}$.
Selectivity, however, depends on the charge of the slit system.
Figure~\ref{fig3} shows the selectivity curves for three smaller membrane charges.
The first panel shows the data for an uncharged slit system.
The charge of the ions has no effect in this case. Therefore, the small ion will enter the slits in a larger quantity (Sm C$^{+}$) along with the counterions (A$^{-}$).

For a small value of the membrane charge ($\sigma_{\mathrm{M}}=-0.05$~C\,m$^{-2}$), the curves are nearly linear.
No selectivity appears to be present in this case.
This is a result, however, of the cancellation of two competing effects.
Its small size is advantageous for the monovalent cation, while its double charge is advantageous for the divalent cation.
At this membrane charge, the two effects balance each other.

Increasing (in magnitude) the membrane charge to $\sigma_{\mathrm{M}}=-0.1$~C\,m$^{-2}$ increases the importance of electrostatic effects and favors the divalent ion.
The selectivity curves are not linear and cross each other at larger $[\mathrm{Sm\, C}^{+}]$ concentrations.
That is, more Sm C$^{+}$ ions are needed to replace the Lg C$^{2+}$ ions in the slits.
Figure~\ref{fig1}~(b) shows even stronger divalent selectivity for $\sigma_{\mathrm{M}}=-0.15$~C\,m$^{-2}$.

The selectivity behavior of this electrolyte mixture, therefore, depends on the membrane charge.
Membrane charge determines which effect will dominate: the small size (entropic advantage) or the large charge (electrostatic advantage) of the ion.
That is why we called this case the ``competitive'' case.

\begin{figure}[ht]
 \centerline{
\includegraphics*[width=0.42\textwidth]{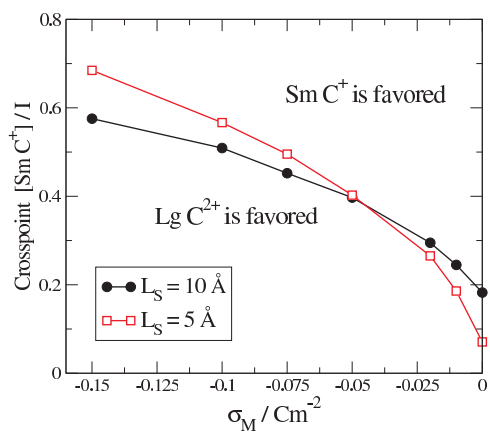}
 }
\caption{(Color online) We define the ``crosspoint'' as the $[\mathrm{Sm\, C}^{+}]/I$ fraction where the occupancies of the two competing cations are the same.
This figure plots the crosspoint as a function of the membrane charge for two values of slit width.
For a given membrane charge, $[\mathrm{Sm\, C}^{+}]/I$ fractions above this curve produce more Sm C$^{+}$ in slit, and less under it.
}
\label{fig4}
\end{figure}

We define the crosspoint as the  $[\mathrm{Sm\, C}^{+}]/I$ value at which the two curves cross each other, namely, at which there are equal numbers of Sm C$^{+}$ and Lg C$^{2+}$ ions in the slits.
We can characterize selectivity with this value.
If it is close to 0, the system is selective for the Sm C$^{+}$ ions.
If it is close to 1, the system is selective for the Lg C$^{2+}$ ions.
Figure~\ref{fig4} shows the dependence of this crosspoint on membrane charge for two different slit widths.
The figure clearly indicates that the divalent ions are favored at large membrane charges.
The effect is more pronounced for narrow slits.

\begin{figure}[!h]
\includegraphics[width=0.48\textwidth]{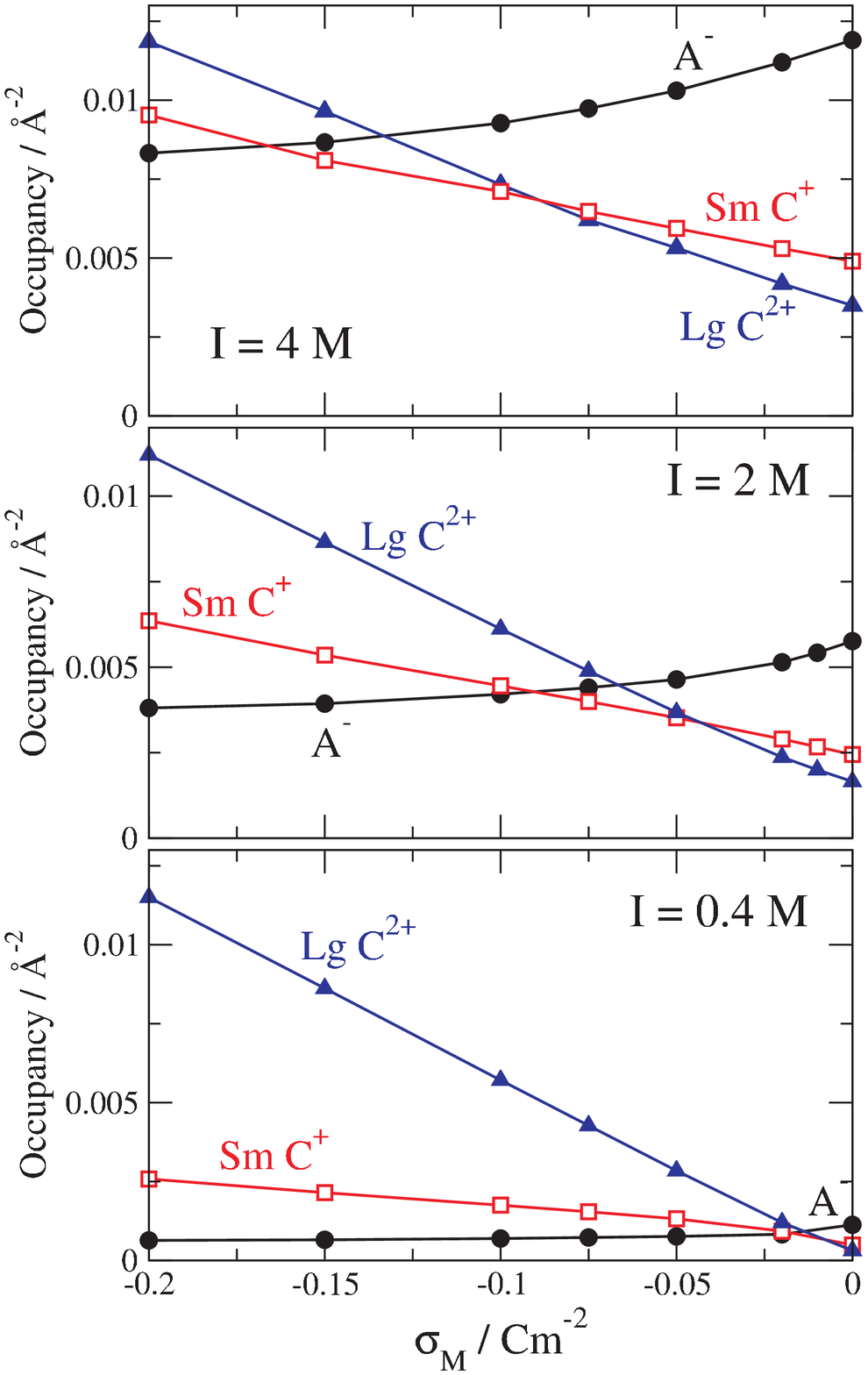}%
\hfill%
\includegraphics[width=0.48\textwidth]{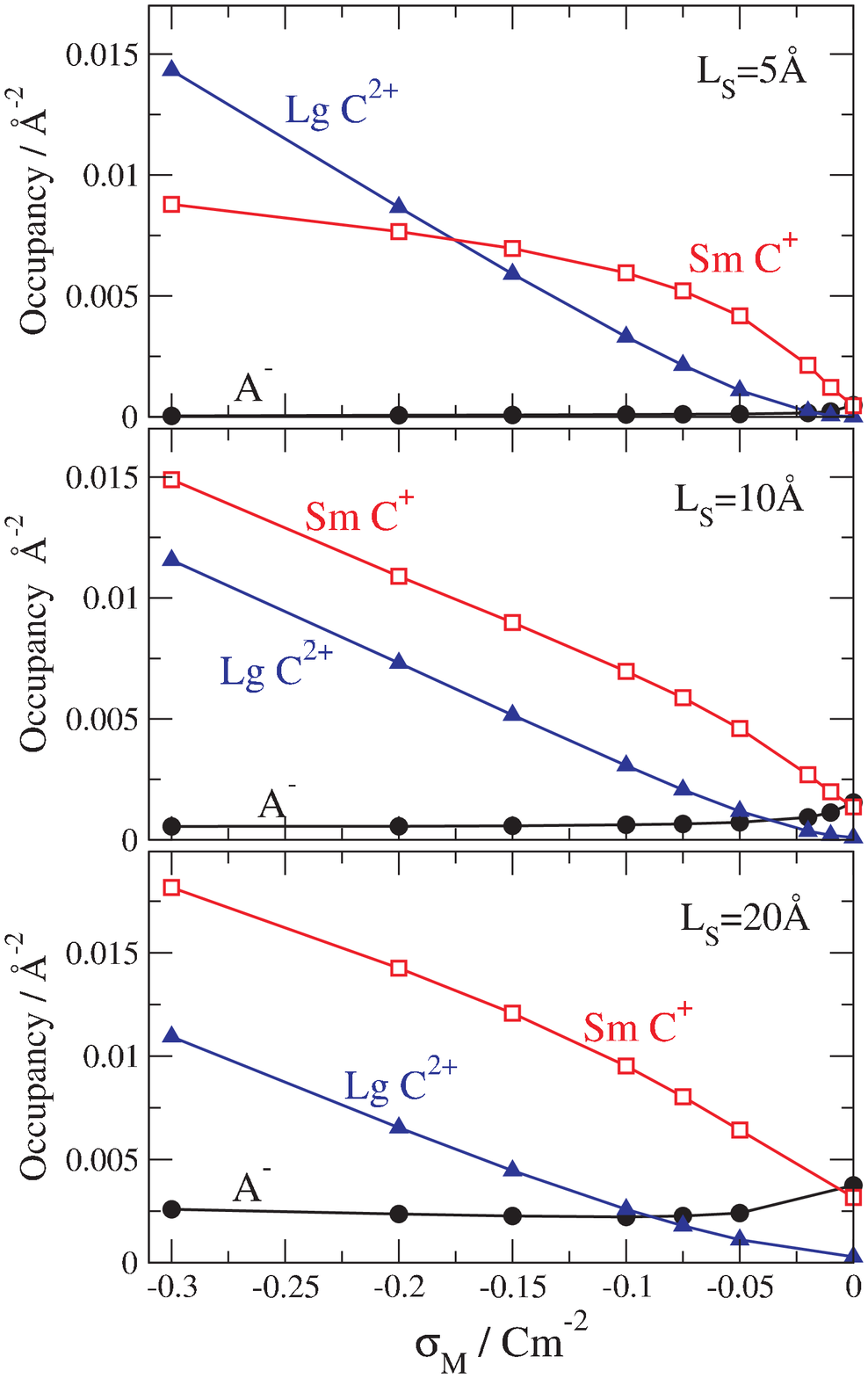}%
\\%
\parbox[t]{0.48\textwidth}{%
\centerline{(a)}%
}%
\hfill%
\parbox[t]{0.48\textwidth}{%
\centerline{(b)}%
}%
\caption{(Color online) Occupancies of various ions for the competition of monovalent and large divalent cations (``competitive'' case) in the central slit ($N_{\mathrm{S}}=3$) as functions of the membrane charge.
(a) Dependence on ionic strength for $L_{\mathrm{S}}=10$~{\AA} and $[\mathrm{Sm\, C}^{+}]/I=0.25$.
Occupancies are shown as functions of $[\mathrm{Sm \, C}^{+}]/I$ with $I=0.4$ M.
For this case, the bulk concentrations of the two cations are the same: $[\mathrm{Sm\, C}^{+}]=[\mathrm{Lg\, C}^{2+}]=I/4$.
(b) Dependence on slit width for $I=0.4$ M and $[\mathrm{C}^{+}]/I=0.75$, which means that the cation concentrations are $[\mathrm{Sm \, C}^{+}]=0.3$ M and $[\mathrm{Lg \, C}^{2+}]=0.0333$ M).
}
\label{fig5}
\end{figure}

This can be plotted in a different way by fixing the bath composition and showing the results as functions of the membrane charge.
This is shown in the top panel of figure~\ref{fig5}~(b).
Sm C$^{+}$ ions dominate at small membrane charges, while Lg C$^{2+}$ dominate at large membrane charges.

Interestingly, this effect is exactly the opposite of what we found at a highly charged electrode \cite{valisko-jpcc-111-15575-2007}.
There, we found (see figure~6 of that paper) that monovalents were favored at very large ($|\sigma|>1$~C\,m$^{-2}$) surface charges.
We did not use such large membrane charges in this study.
The space in a slit is limited; at this charge there would not be enough space for the cations to balance the membrane charges inside the slits.
This being not possible, they would balance the charge of the slit system from outside, from the outer DLs.
That practically would be the case considered in our previous work for the isolated interface \cite{valisko-jpcc-111-15575-2007}.
The explanation of our earlier result that small monovalent ions dominate near the highly charged surface is that the density is so large close to the surface that it favors the small ions.
There is, however, enough space in the diffuse layer for the divalent ions farther from the electrode so they can contribute to screening from afar.

Figure~\ref{fig5} plots the selectivity curves as functions of the electrode charge.
Figure~\ref{fig5}~(a) analyzes the effect of the total ionic strength, while figure~\ref{fig5}~(b)  analyzes the effect of the slit width.
As figure~\ref{fig5}~(a) shows, there is a much stronger competition between the two cations at a small ionic strength.
No anions are adsorbed into the slits and the slits must decide which ion they like at the given composition and membrane charge (bottom panel).
Increasing the ionic strength, the selectivity becomes less dependent on membrane charge.
Both cations are adsorbed in the slits and their relative occupancy does not really depend on $\sigma_{\mathrm{M}}$.
Interestingly, the quantity of anions depends on $\sigma_{\mathrm{M}}$ sensitively: increasing $\sigma_{\mathrm{M}}$, anions are repelled from the slits.

Figure~\ref{fig5}~(b) shows the occupancy curves for different slit widths.
As expected, there is a stronger competition between the Sm C$^{+}$ and Lg C$^{2+}$ ions in the narrow slit (top panel).
Smaller membrane charge favors the small cations, because their quantity is larger in the bulk ($[\mathrm{Sm\, C}^{+}]/I=0.75$).
Increasing the membrane charge, the divalent cations gain advantage from the strong competition for space in the slits.
At larger slit widths, selectivity is less sensitive to the membrane charge.
There are always more Sm C$^{+}$ cations in the slits in accordance with the larger bulk concentration of these cations ($[\mathrm{Sm\, C}^{+}]/I=0.75$).
The relative amount of the two competing cations depends on slit width; the narrower the slit is, the more selective it is for the divalent ion.

Selectivity for the divalent ion against the monovalent ion was also observed by Jamnik and Vlachy \cite{jamnik_jacs_1995}, who simulated the competitive ion partitioning between a charged micropore and the bulk.
The radius of the pore in that study, however, was quite large (40 \AA), so the DLs did not overlap.
Divalent vs. monovalent selectivity is weaker in this case as confirmed by figure~\ref{fig5}~(b).

\begin{figure}[h]
\centerline{
\includegraphics*[width=0.42\textwidth]{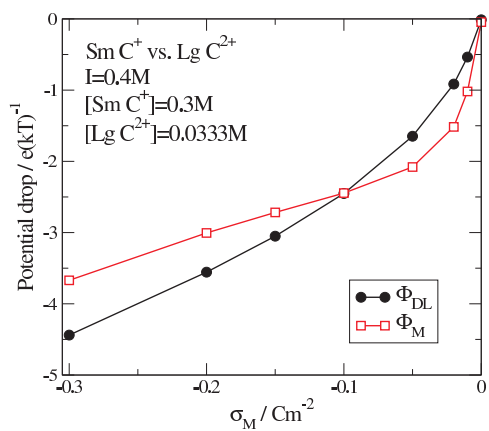}
}
\caption{(Color online) Figure \ref{fig2} defines the membrane potential, $\Phi_{\mathrm{M}}$, and the double layer potential, $\Phi_{\mathrm{DL}}$.
The membrane potential is the potential difference between the bulk and the inner membranes.
The DL potential is the potential drop across the outer DLs.
 This figure plots these potential drops as functions of the membrane charge for the ``competitive'' case with $I=0.4$ M and  $[\mathrm{Sm\, C}^{+}]/I=0.75$.
}
\label{fig6}
\end{figure}

Finally, we return to the bottom panel of figure~\ref{fig2}~(b), where the mean potential curves are plotted for different membrane charges.
This figure also defines the potential differences $\Phi_{\mathrm{M}}$ and $\Phi_{\mathrm{DL}}$.
These potential values are shown as functions of the membrane charge in figure~\ref{fig6}.
The different shapes of the potential curves in figure~\ref{fig2} have the effect on the curves in figure~\ref{fig6}.
It is seen that $|\Phi_{\mathrm{M}}|>|\Phi_{\mathrm{DL}}|$ at small membrane charges (in absolute value), while the reverse is true for large membrane charges.

The explanation of this behavior has been given at the discussion of figure~\ref{fig2}~(b).
The basis is that we ``overcharge'' the outer slits at large membrane charges.
This is because the large membrane charge favors the divalent ions in the slits.
The DLs outside the slit system, on the other hand, is less selective for the divalent ion because strong competition is not enforced by a limited space.
The excess of divalents in the outer slits, then, ``overcharges'' these slits.
According to charge neutrality, there is a shortage of ionic charge in the outer DLs.

Our simulations for selective adsorption of various competing cations in charged slits revealed that the basic behavior is the same as in the case of calcium channels: the narrow slits favor small ions and divalent ions.
The interesting case of the ``competitive'' case (Sm C$^{+}$ vs. Lg C$^{2+}$) was discussed  in more detail.

% \section{Conclusions}
% \label{sec:conclusions}

\section*{Acknowledgements}

We acknowledge the support of the Hungarian National Research Fund (OTKA K68641) and the J\'anos Bolyai
Research Fellowship.
Present publication was realized with the support of the project T\'AMOP--4.2.2/A--11/1/KONV--2012--0071 and T\'AMOP--4.1.1/C--12/1/KONV--2012--0017.
We are grateful for a generous allotment of computing time at the MARYLOU supercomputing facility of Brigham Young University.

\clearpage

\ukrainianpart

\title{Селективна адсорбція іонів у заряджених системах між двома
площинами}

\author{М. Валіско\refaddr{label1}, Д. Гендерсон\refaddr{label2}, Д. Бода\refaddr{label1,label2}
}

\addresses{
 \addr{label1} Факультет фізичної хімії, університет Паннонії, Веспрем, Угорщина
 \addr{label2} Відділ хімії та біохімії,  Університет Брайхем Янг, Прово, Юта, США
}

\makeukrtitle
\begin{abstract}
Ми досліджуємо селективну адсорбцію різних катіонів у шарувату
щілиноподібну систему, використовуючи  Монте Карло симуляції у
великому канонічному ансамблі. Щілиноподібна система формується
послідовністю негативно заряджених мембран. Електроліт містить два
типи катіонів різних розмірів і валентностей, які моделюються
зарядженими твердими сферами, зануреними в однорідний діелектричний
розчинник. Ми представляємо результати для різних випадків в
залежності від комбінації властивостей конкуруючих катіонів. Ми
розглядаємо випадок, коли двовалентні катіони є більші, ніж
моновалентні. У цьому випадку, розмір і заряд мають
зрівноважувальний ефект, що приводить до цікавих селективних явищ.

\keywords Монте Карло, примітивні моделі електролітів, щілини,
селективність

\end{abstract}

\end{document}